
\input harvmac
\def\L{\Lambda}
\def\CO{{\cal O}}
\def\l{\ell}
\def\CA{{\cal A}}
\def\W{\hat W}
\def\V{\hat V}
\def\v{\hat v}
\def\w{\hat w}

\def\tw{\tilde w}
\def\G{\Gamma}
\def\tl{\tilde \l}
\def\tt{ T}
\def\t {\tilde}
\def\CC{ {\cal C}}
\def\CV{{\cal V}}
\def\e{\epsilon}
\def\gab{ G_{ab}}
\def\p{\partial}
\def\dxi{d^{2}\xi \sqrt{{\rm det}G(\xi)}}

\def\gst{\gamma _{str}}
\def\CD{{\cal D}}
\def\R{\relax{\rm I\kern-.18em R}}
\font\cmss=cmss10 \font\cmsss=cmss10 at 7pt
\def\Z{\relax\ifmmode\mathchoice
{\hbox{\cmss Z\kern-.4em Z}}{\hbox{\cmss Z\kern-.4em Z}}
{\lower.9pt\hbox{\cmsss Z\kern-.4em Z}}
{\lower1.2pt\hbox{\cmsss Z\kern-.4em Z}}\else{\cmss Z\kern-.4em Z}\fi}
\def\pl{{\it  Phys. Lett.}}
\def\np{{\it Nucl. Phys. B}}

\def\s{\sigma}

\def\i{{\rm Im}}
\def\intoi{\int _{0}^{\infty}}
\def\o{ 2 \cos (\pi p_{0}/2)}
\def\xo{\sqrt{ 2 \cos (\pi p_{0}/2)}}
\def\CM{{\cal M}}
\def\dbm{ \p {\cal M} ^{(D)}}
\def\dnm{ \p {\cal M} ^{(N)}}
\Title{\vbox{\baselineskip12pt\hbox{RU-92-10}\hbox{LPTENS-92/11}}}
{Loop Gas Model for Open Strings}
\centerline{V. A. Kazakov}
\bigskip\centerline{Department of Physics an Astronomy}
\centerline{
Rutgers University, Piscataway, NJ 08855-0849, USA}
\centerline{and}
\centerline{Laboratoire de Physique Th\'eorique}
\centerline{Departement de Physique de l'Ecole Normale Sup\'erieure}
\centerline{ 24 rue Lhomond, 75231 Paris Cedex 05, France}

\bigskip\centerline{ I. K. Kostov \footnote{$^\ast $}
{ on leave from the Institute for Nuclear Research and Nuclear Energy,
72 Boulevard Tsarigradsko Shosse, 1784 Sofia, Bulgaria}}
\bigskip\centerline{Service de Physique Th\'eorique  \footnote{$^{\dagger}$}{
Laboratoire de la Direction
 des Sciences de la Mati\`ere du
Commissariat \`a l'Energie Atomique} de Saclay }
\centerline{CE-Saclay, F-91191 Gif-sur-Yvette, France}


\vskip .3in

\baselineskip8pt{
The open string with one-dimensional target space is formulated in terms
of an SOS, or loop gas,  model on a random surface.
We solve an integral equation for the loop amplitude with
Dirichlet and Neumann  boundary conditions
 imposed on  different pieces of its boundary.
The result is used to calculate the mean values of order
 and disorder operators,  to construct the string propagator and find its
 spectrum of excitations. The latter is not sensible neither to the
string tension $\L$ nor  to
the mass $\mu$ of the ``quarks'' at the ends of the string.
As in the case of closed strings, the SOS formulation allows to
construct a Feynman diagram technique for the string
interaction amplitudes.
\smallskip
\leftline{Submitted for publication to: {\it Nuclear Physics B}}
\rightline{SPhT/92-049}
\Date{4/91} 


\newsec{Introduction}
The open string theories always attracted a considerable attention of
the physicists, not only from the point of view of critical strings but
also as a possible source of field theoretical applications. For
example,
the idea to formulate the multicolor QCD as a theory of noninteracting
strings (random surfaces) has been fascinating the minds of some
theoretical physicists in the 80's.
This string theory   should involve  both closed strings
describing glueballs and open strings describing the $q \bar q$  bound
states (mesons).

Now, after more than ten years of  study, we know how to formulate and
solve the simplest theory of random surfaces - the (noncritical)
 closed bosonic string.
 In order to go further, an obligatory exercise to do is to
extend the solution to the case of open strings.

It is clear that the physics of open strings should be more complicated
 than that of
closed strings, since it depends on the choice of the boundary conditions
at the ends of the string. The  string  amplitudes will depend now on two
dimensionful parameters: the string tension  $\L$  coupled
to the area of the world sheet
(the ``bulk cosmological constant'' in the language of 2d gravity)
  and the mass $\mu$ at the ends of the string
 coupled to the length of the boundary
of the world surface (the ``boundary cosmological constant'').

The open bosonic string is well defined for embedding spaces with
effective dimension (``the central charge of the matter fields'') $
-\infty  <C \le 1$;
otherwise the vacuum would be unstable due to the tachyonic excitation.
The field theory of open strings with no embedding ($C=0$) has been
formulated as a random matrix model in
\ref\kaz{V. Kazakov, \pl , 237B(1990) 212}. This model was
then solved in the double scaling limit in
\ref\kos{I. Kostov, \pl , 238B(1990) 181} (see also
\ref\yy{Yukihisa Itoh and Yoshiaki Tanii, preprint STUPP-92-127,
February 1992}). Further, the theory of $C=1$
open strings (embedding space $\R$) was considered as a solution of
matrix quantum mechanics in \ref\affl{I. Affleck, \np \ 185 (1981) 346 },
\ref\mike{M.Douglas, to be published}
and \ref\minahan{J.Minahan, preprint UVA-HET-92-01, March 1992}.
Open strings with $C=-2 $ and $C=1$   have been also considered in
\ref\bula{D. Bulatov, {\it Int. Journ. Mod. Phys. } A6(1991) 79  }.

On the other hand, the noncritical open strings have been studied
by means of the Liouville theory
\ref\Bersh{M.Bershadsky and D.Kutasov, preprint PUPT-1283, HUTP-91/A047
(1991)}, \ref\Jas{Z.Jaskolski, ICTP-preprint (1991)},
\ref\Mar{E.Martinec, G.Moore, andN.Seiberg, preprint RU-14-91,
YCTP-P10-91, EFI-91-14 (1991)}, \ref\Tan{Y.Tanii and S.Yamaguchi,
preprint STUPP-91-128 (1992), STUPP-91-121 (1992)}.
The continuum approach is based on a free field theory (the Liouville
potential is treated as a perturbation) and  therefore
cannot be used to evaluate
the full string interaction  amplitudes.
This approach  is sufficient to study the so-called bulk amplitudes
which obey the conservation of the Liouville energy.

 In this paper we propose a systematic approach to the open
noncritical strings with $-\infty <C \le 1$ which can be used to
find the exact string interaction amplitudes.
A very convenient framework for this purpose is provided by the loop gas
(or SOS-, or heights-) model on a random surface
\ref\Iade{I. Kostov, \np 326(1989)583}
\ref\Icar{I. Kostov, ``Strings embedded in Dynkin diagrams'',
lecture given at the Cargese Workshop on Random Surfaces;
Saclay preprint SPhT/90-133}
\ref\Inonr{I. Kostov, \pl B266(1991) 312}
\ref\Imult{I. Kostov, \pl  B266(1991)42}
\ref\Idis{I. Kostov, ``Strings with discrete target space'',
Saclay preprint SPhT/91-142, to appear in \np }.
The target space in this model is the infinite discretized line $\Z$.
It is  sometimes called loop gas model because the domain walls
between the regions of constant height form a configuration of
nonintersecting loops on the world sheet. In the case of  Dirichlet boundary
condition the points along a connected  boundary have the same height
and therefore the domain walls cannot end at the boundary.

Below we are going to adapt this model for the case of Neumann boundary
conditions corresponding to free endpoints of the open string.
In this case the domain walls are always orthogonal to the edge of the
world sheet. This means that the loops are repulsed from the boundary but
domain walls can approach it at right angle.  We will solve the loop equation
for the amplitude of a disk with a boundary divided into two parts
with Dirichlet and Neumann boundary conditions respectively.
Knowing this amplitude we can further calculate the open string propagator
and the string interactions following the strategy applied in the case
of the closed SOS string \Imult , \Idis .
The eigenstates diagonalizing the string propagator are different for
different choices of the parameters $\L$ and $\mu$ of the open string.
However, the diagonalized propagator is universal and is in fact identical
to the one of the closed SOS string.

In order to explore the whole range of effective dimensions of the
target space $- \infty <C\le 1$ we will introduce, following the
Coulomb gas picture,  a distributed  background momentum
(``electric charge'')
 proportional to the
curvature of the world sheet metric.
The momentum conservation (the electric charge neutrality)
is assured by introducing
 pointlike electric charges at the critical points of the
embedding of the world sheet.  On the lattice this construction has
been elaborated in \Idis \ and \ref\nt{O. Foda and B. Nienhuis, \np
324(1989) 643}.

In the present paper we avoid introducing lattice discretization of
the world sheet in order to keep closer connection with the continuum
theory. A derivation of the basic loop equation  for a discretized surface
is presented in the Appendix.


\newsec{Definition of the model}
{\it 2.1. Coulomb gas picture}
\smallskip
The dynamical fields in the Polyakov formulation of the string path
integral \ref\polyak{A.M. Polyakov, \pl B103(1981) 207}
are the position field $x(\xi )$ and the intrinsic metric $\gab (\xi) ,
a,b=1,2$ of the world sheet (which we will denote by $\CM$).
We will assume that $\CM$ has  the topology of a disk. The boundary $\p \CM$
is divided into $2n$ pieces as is shown in \fig\frstt{The geometry
of the world sheet with alternative Dirichlet-Neumann boundary
conditions. The thick lines represent the Neumann boundaries}
 on which we impose alternatively
 boundary conditions of Dirichlet
\eqn\diri{\p _{\|} x(\xi)\equiv t^{a}(\xi)\p _{a}x(\xi)=0, \ \ t_{a}(\xi)=
\ {\rm unit \ tangent \ vector}}
 and
Neumann
\eqn\neum{\p _{\bot} x(\xi) \equiv n^{a}(\xi)\p_{a} x(\xi)=0, \ \ n_{a}(\xi)=
\ {\rm unit \ normal \ vector}}
We will denote the Dirichlet boundary by $\p \CM ^{(D)}$ and the
Neumann boundary by $\p \CM ^{(N)}$.
Each kind of boundary consists of $n$ connected pieces
\eqn\pisec{\eqalign{
\dbm& = \dbm_{1}+...+\dbm_{n},\cr
\dnm& = \dnm_{1}+...+\dnm_{n};\cr
\p \CM& = \dbm + \dnm \cr}}

 The Dirichlet boundary condition \diri\
is appropriate for the initial anf final string
states; it describes a boundary which occupies
a single point of the embedding space. The Neumann boundary condition
means that the flow of energy across the boundary is zero; it should be imposed
along the edges of the world sheet representing the endpoints of the open
string.

The world sheet with $n$ pairs of boundaries describes the
interaction of $n$ open strings.
The corresponding  amplitude will depend on the intrinsic geometry of the
world sheet only through the gauge invariant quantities: the total
area of the world sheet
\eqn\totar{A=\int _{\CM}dA(\xi); \ \ \  \ dA(\xi)= \dxi  }
and the total lengths of the Dirichlet and Neumann  boundaries
\eqn\totdl{\l_{k} = \int _{\dbm _{k} } d \l (\xi) , \ \ \tilde
 \l_{k} = \int _{\dnm_{k}}
d\l (\xi), \ \  k=1,...,n; \  \ \  \  d\l (\xi)= t_{a}(\xi)d \xi ^{a}}

   An effective dimension $C<1$ can be achieved
by introducing a coupling $p_{0}$ (``distributed electric charge'' in the
Coulomb gas terminology) between the field $x$ and the intrinsic
geometry of the world sheet.
 The world sheet action  then reads
\eqn\pact{\eqalign{
\CA [x, \gab ] =& \CA '[x, \gab ] +\CA '' [x, \gab ] \cr
\CA '[x, \gab ]  =& {g \over 4\pi } \int _{\CM}
dA(\xi)  G ^{ab}
\p _{a}  x(\xi) \p _{b} x(\xi)\cr
 \CA ''[x, \gab ] =
 & ip_{0}\Big[{1 \over 4\pi} \int _{\CM} dA(\xi) x(\xi)\hat R(\xi)
+ {1\over 2\pi} \int _{\p \CM} d\l(\xi) x(\xi) \hat K(\xi)\Big]\cr}}
where   $\hat R(\xi)$ is the intrinsic Gaussian curvatire at the point $\xi
\in \CM$ and $\hat K(\xi) $ is the geodesic curvature at the point
$\xi \in \p \CM$. The factor  $g$  known as
the Coulomb gas coupling constant can be elliminated by
rescaling $x$. We fix the normalization of $x$ to have
\eqn\xpxpxp{g=1+p_{0}.}
The two curvatures are  normalized so that the Gauss-Bonnet formula reads
\eqn\gbfor{\int _{\CM}
 d^{2}\xi \sqrt{\det G}  \hat R(\xi )  + 2 \int _{\p \CM} d\l (\xi)
\hat K(\xi) = 4 \pi }
The boundary term in \pact\ is introduced in order to  be able to
satisfy the momentum conservation  (the ``electric charge neutrality'').
It is clear from the
  Gauss-Bonnet formula \gbfor\ that   the zero mode $ x(\xi)= x_{0}$
 of the $x$-field
 produces only a factor $\exp (-i p_{0}x_{0})$ and can be neutralized by
introducing a background momentum $-p_{0}$ at some point of the
boundary.

Eq. \pact\ defines the standard Coulomb gas description of the $C \le 1$
strings. In this paper we propose a modified version of the Coulomb gas
approach in which the electric charge neutrality is required in a
stronger sense. We introduce a system of pointlike electric charges
associated with the points where the string picture changes.
These are the critical points of the map $\CM \to \R$
\eqn\crtpnt{dx(\xi)= \p_{a} x d \xi ^{a} =0}
shown in \fig\dve{Critical points of the embedding of the world sheet.
a) Creation (annihilation) of a closed string state.
b) Splitting (joining) of closed strings.
c) Creation (annihilation) of open string state.
d) Splitting (joining) of open strings.}.
We distinguish four kinds of critical points
$\xi ^{\ast}$  which will be characterized by
a weight $\chi (\xi ^{\ast})$  taking values $1, -1, 1/2, -1/2$.

For the   critical points in the interiour of the world sheet
(cases $a$ , $b$ ) we define
\eqn\iaiaia{\chi (\xi ^{\ast})= {\rm sgn \ det} \|\p _{a}\p_{b}
 x(\xi) \|  _{\xi = \xi ^{\ast} } , \ \ \ \xi ^{\ast} \in \CM}
For the critical points along the edge of the world sheet
(cases $c$ , $d$ )  we define
\eqn\ibibib{\chi (\xi ^{\ast})= {1 \over 2} {\rm sgn}\  \p _{\|}^{2}x(\xi)_{\xi
 = \xi ^{\ast}} }
The sum over the weights of all critical points gives the
Euler characteristics of the world sheet
\eqn\eulll{\sum_{ \xi ^{\ast}} \chi (\xi ^{\ast}) \ = \ \chi }
Therefore, if we associate with each critical point  $\xi ^{\ast}$
a charge $ -p_{0} \chi( \xi ^{\ast})$, the electric charge neutrality
will be fulfilled.
The factor
\eqn\csssn{\prod _{\xi ^{\ast}} e^{-ip_{0}\chi (\xi ^{\ast})}}
can be taken into account by adding
 to the standard action \pact\ a second linear  term
\foot{Strictly speaking, this term is not linear in $x$ because of
the charge density $\rho (\xi)$ depending on the embedding $\xi \to x(\xi )$.
Note also that each connected Dirichlet boundary contributes a factor
$-p_{0}$ if it is a closed loop and $-p_{0}/2$ if  it is an open interval.}
\eqn\hhhatn{ \CA ''' = -p_{0}  \int _{\CM}d^{2}\xi  \ x(\xi)
\rho (\xi); \ \rho (\xi) = \sum _{\xi ^{\ast}} \chi (\xi^{\ast})
\delta _{\xi , \xi ^{\ast}}}
The density $\rho (\xi)$ can be expressed
through the vector field with unit norm $\hat n (\xi)$
\eqn\hatn{ {\rho (\xi)
 =   {1 \over 2\pi}
\int _{\CM}d^{2}\xi  x(\xi) \varepsilon  ^{ab} \varepsilon  ^{cd}
  \p_{a} \hat n _{c} \p _{b} \hat n _{d}}}
\eqn\normpr{\hat n _{a} (\xi) = {\p _{a} x(\xi) \over \sqrt{\p _{a}
x(\xi) \p ^{a} x(\xi)}}}
Consider  the  functional integral
\eqn\fint{ Z(A; \tl _{i},\l _{i}, x_{i}, i=1,2,...,n)
=\int [dx(\xi)][dG_{ab}(\xi)] e^{-  \CA[x,G_{ab}]}}
\eqn\hhatn{ \CA[x, \gab]=  \CA ' + \CA '' + \CA '''}
where  the integral over intrinsic geometries is restricted to
surfaces with fixed area $A$ and lengths $\l _{i}$ and $\tl _{i}$
of the Dirichlet and Neumann boundaries, correspondingly.
The integral depends also on the positions $x_{i}$ of the Dirichlet
boundaries in the embedding space.
The  interaction amplitude of $n$ open strings with momenta
$p_{1},...,p_{n}$ and lengths $\l_{1},...,\l_{n}$
is defined by the Laplace transform
\eqn\defam{\eqalign{
& v( p_{k},
\l _{k};   k=1,...,n)  = \cr
&\int _{0}^{\infty}dA e^{-A\L} \prod _{k=1}^{n}
\int _{0}^{\infty}d\tl _{k} e^{ - \mu \tl_{k} +ip_{k}x_{k}}
 Z(A; \tl _{i},\l _{i}, x_{i}, i=1,2,...,n) \cr}}
Here $x_{k}$ denotes the position of the $k$-th Dirichlet boundary
$\dbm_{k}$. The string tension $\L$ coupled to the  area of the
world sheet is called sometimes cosmological constant, since this functional
integral can be also considered as the partition function for two-dimensional
quantum gravity. Similarly, the mass $\mu$ of the ends of the
string can be also interpreted as a boundary cosmological constant since it
is coupled to the length of the Neumann boundary.

The amplitude \defam\ is nonzero only if
 the sum of all momenta
 is equal to the
background momentum $ (1-n/2) p_{0}$
\eqn\consvtn{p_{1}+...+p_{n}=(1-{n \over 2} )p_{0}}
It is convenient to introduce the variables $z_{k}$
dual to
the lengths $\l_{k}$ and consider the Laplace image of \defam\
\eqn\lapam{\v (p_{1},z_{1};...;p_{n},z_{n})
= \int _{0}^{\infty} d\l_{1}  ... \int _{0}^{\infty} d\l_{n}
\ \ e^{-\l_{1}z_{1}-...-\l_{n}z_{n}} v(p_{1},\l_{1};...;p_{n},\l_{n})}

The presence of the background momentum $p_{0}$ diminishes the
effective dimension of the embedding space (the conformal anomaly
due to the matter  field) from one  to
\eqn\centrch{C=1-6 p_{0}^{2}/g =1-6(g-1)^{2}/g}
and restricts the spectrum of allowed momenta to
\eqn\spctrmp{p= {k \over 2}  p_{0},\ \ \  \ k \in \Z}

The local operators in the theory are those creating microscopic closed
and open strings. Sometimes they are called bulk and boundary operators\Mar .
The spectrum of the bulk operators can be fixed with self-consistency
arguments
known as  David-Distler-Kawai analysis
\ref\ddk{F. David, {\it Mod. Phys. Lett.} A3(1988) 1651; J. Distler and
H. Kawai, \np 321(1989) 509},  based on the
assumption that at distances large compared to the cutoff
 but small compared to the
 size of the world sheet,
 the fluctuations of the metric $\gab$ are described
  by a gaussian field.

Below we  present a sketch of
 these arguments mainly to help the reader to become familiar with our
normalization which is not the standard one used in the string theory.

After introducing a conformal gauge
\eqn\cong{G_{ab} = e^{2\nu \phi (\xi)} \gab ^{0}(\xi)}
where $\gab ^{0}$ is some fiducial metric, and taking account of the
conformal anomaly we arrive at an effective action depending on a
two-component gaussian field $(x,\phi)$:
\eqn\efact{\eqalign{
&  \CA _{{\rm Liouville}} [x,\phi] = {1 \over 4 \pi } \int  d\xi
 \sqrt{ {\rm det}\hat G^{0}(\xi)} [ g \ G ^{0ab}
(\p _{a}  x(\xi) \p _{b} x(\xi)\cr
& -\p_{a}\phi (\xi) \p_{b}\phi (\xi ) )
 + ( ip_{0} \ x(\xi) - \epsilon _{0} \phi (\xi))
\hat R^{0}(\xi) ]\cr}}
The vertex operator creating a momentum $p$, dressed by the fluctuations of the
metric is\foot{Here we write explicitely the compensating charge $-p_{0}$
associated with the puncture}***
\eqn\vertop{{\CV_{(p,\e)}(\xi)= e^{i(p-p_{0})x(\xi)}e^{-(\e(p)-\e _{0})
\phi (\xi)}}}
In particular, the puncture operator ${\cal P} = - \p / \p \L$ which marks a
point on a surface
 is represented by
\eqn\punct{{\cal P}(\xi) = e^{- (\e(p_{0})-\e_{0})\phi (\xi)}
=e^{2\nu \phi (\xi)}}
The condition of the absence of conformal anomaly yields
\eqn\anom{C_{tot}=C_{(x)}+C_{(\phi)} -26 \equiv 1-6{p_{0}^2 \over g}
+1 +6{\e _{0}^{2} \over g} -26 =0  \ \ \Rightarrow
\e _{0}^{2}-p_{0}^{2}=4g }
We choose the positive solution thus fixing a positive direction in the $\phi$-
space
\eqn\epso{\e _{0}=g+1, \ \ p_{0}=g-1}
The condition that the conformal dimension of the operator \vertop\
is one
\eqn\cdvo{\Delta _{x}+\Delta _{\phi} \equiv {p^{2}-p_{0}^{2} \over 4g} -
{\e(p)^{2} -\e_{0}^{2} \over 4g}=1}
combined with \anom\ leads to the relation
\eqn\epep{\e(p)^{2}-p^{2}=0}
which can be interpreted as a mass-shell condition for the 2-momentum
$(p, \e)$. All physical operators correspond to positive Liouville energies
\eqn\lien{\e(p)=|p|}
The Liouville charge of the identity operator equals to $\e _{0}-\e (p_{0})
=2\nu$ where
\eqn\nuuu{\nu = {1 \over 2}(g+1 - |g-1|)}
  The gravitational dimensions of the vertex operators coherent with the
background momentum $p_{0}=|g-1|$ are
\eqn\grdm{\delta _{rs}= 1- {\e_{0}-\e (p_{rs}) \over \e _{0}- \e (p_{0})}
= {|r-gs|-|g-1| \over g+1 - |g-1|}}
Finally, the string susceptibility exponent  $\gst$ giving the dimension of
the string interaction constant is equal to
\eqn\gstring{\gst = - {\e (p_{0}) \over \nu} =
- {2|g-1| \over g+1 +|g-1|} ; \ \ \ \  \nu (2-\gst )=\e _{0}}

The above arguments can be easily generalized to the boundary operators \Mar .
However, as it has been noticed in \Icar -\Idis , the semiclassical
analysis is not always applicable at the boundary.

\bigskip
{\it 2.2. Formulation as an SOS model on the world sheet}
\smallskip
Let us now try to find a link between this continuous formulation of the
path integral and the so called SOS model in which the $x$ field is
restricted to take only discrete values (heights) $x / \pi  \in \Z$.
At large distances the configurations of such field should look as continuous;
this is achieved by the condition that the $x$ field can jump only with a step
$ \pm \pi $.
The domain walls separating the domains on the world surface where $x$
takes constant value form a pattern of nonintersecting lines. In the
case of a surface without boundary all these lines should be closed loops.
Across each domain wall the heigth $x$ jumps by $\pi$
: $x \rightarrow x \pm \pi $.
The sign can be taken into account by assigning an orientation to the
domain wall.

If we consider a Dirichlet boundary, the above picture holds
unchanged. Since the height $x$ is not changing along the boundary,
the whole boundary belongs to a single domain.
 Note however  that the loops are allowed to touch the boundary and
this should be taken into account when writing the loop equations
for the closed string \Iade .

On the other hand, in presence of a Neumann boundary
 the above geometric picture changes drastically.
The domain walls are not only loops but also lines ending at the boundary
(\fig\vtora{A configuration of domain walls for a world sheet with
Dirichlet-Neumann boundary conditions}).
The condition that the  normal derivative of the $x$ field
is zero
in the vicinity of the boundary means that all these lines meet the
boundary at right angle. In addition, the closed loops are not allowed to
approach the boundary.

The integration  over the $x$ field can be replaced by
a sum over all loop configurations on the world sheet
and a subsequent sum over all allowed values of $x$ in the domains
bounded by these loops:
\eqn\measure{\int \prod _{\xi} dx(\xi) ... \longrightarrow
\sum _{{\rm loop\ configurations}}  \ \ \ \ \
 \  \sum _{x  ({\rm domains})} ... }

Suppose that the integral over the world-sheet intrinsic geometries
is regularized, say, by a discretization using planar graphs.
Then \measure\ can serve as a microscopic definition of the string
path integral.

 Let us ``derive'' the Boltzmann weights of
the domain-wall configurations from the continuum action \hhatn .
We can imagine that the SOS configuration is regularized so that the map
$\CM \to  \pi \Z$ is obtained as a limit of a smooth map $\CM \to \R$.
Then the contribution of the last term in \hhatn\ comes only from the
vicinity of the domain walls.
Notice that    $d\varphi (\xi)  = \hat n (\xi)  \times  \p_{a} \hat n
(\xi)d\xi^{a} $ is the infinitesimal angle
swept by the unit vector $n_{a}(\xi)$  along the interval $ d \xi$.

Let us  consider an SOS configuration
of the field $x(\xi)$ described by a system of domains and domain walls
and evaluate the action \hhatn .

The term $\p _{a} x \p ^{a} x$  in the integrand
 is just a square of the invariant
gradient of the $x$-field, which is zero everywhere except of the
domain walls, where it is an (infinite) positive constant along the wall. Being
integrated over the world surface it yields
the total length of the domain
 walls times a (cut-off dependent) positive factor.
Thus its contribution to the action is
\eqn\actn{\CA' =  K_{0} \int _{{\rm domain \ walls}} d\l(\xi) =
K_{0} \l _{\rm total}}
where  $d\l(\xi)$ is the length element
along a domain wall and  $ \l _{\rm total}$ is the total length of the domain
 walls on the world sheet.

tNow let us demonstrate that
he contribution of  the last two terms in \pact\
  depends {\it only} on the topology of the
configuration of domains and domain walls.
 Let us consider a world
sheet with the topology of a disk and a system of
domain walls separating the domains $\CD_{1}, \CD_{2}, ...,
\CD_{k},... $ on it. The domain $\CD_{k}$  is bounded by domain walls
(loops and open lines) and pieces of the boundary of the world sheet.
Two neighbour domains can be separated either by a closed loop or by an
open line.
For each two domains having a common boundary one of them
is surrounded by the other; therefore the system of domains has a tree-like
structure.
The boundary of  each domain $\CD_{k}$
consists of  $c_{k}$  connected components. The $c_{k}-1$ internal boundaries
are all closed loops. The  externel boundary
is  made out of   $n_{k}$ open lines separated by pieces of the boundary of the
world sheet. $n_{k}=0$ means that the external boundary is also a closed loop.

Now we can express the contribution of the term in \pact\ proportional to the
 gaussian curvature in terms of  the
heights $x_{k}$ of the domains   $\CD _{k}$ and the numbers $c_{k},n_{k}$
characterizing the topology of the domain wall configuration.
Applying the Gauss-Bonnet formula to each domain domain $\CD_{k}$ we find
\eqn\gb{\eqalign{
& \CA '' -i{p_{0} \over 2\pi} \int _{\p \CM}d\l (\xi) x(\xi) \hat K(\xi) \cr
& =i {p_{0} \over 4\pi}\int _{\CM} dA(\xi) x(\xi) \hat R(\xi )
 \cr
&= i {p_{0}\over 4\pi}
\sum _{k} x_{k} \int _{\CD_{k}} dA(\xi) \hat R(\xi ) \cr
 &= - i{p_{0}\over 4\pi}  \sum _{k} x_{k}
\Big[2\Big( \int _{\p \CD_{k}}d\l (\xi)
\hat K(\xi) +  \pi n_{k}\Big)+  4 \pi ( h_{k} +c_{k} -2)\Big]\cr}}
where $\hat K(\xi) d\l(\xi)$ is the infinitesimal angle swept by the normal
vector $n_{a}(\xi)$ along the boundary $\p \CD_{k}$, and
 $h_{k}$ denotes the enclosed genus (\# handles) in the domain
$\CD _{k}$.  Since we are considering the topology of the disk, $h_{k}=0$.
The boundary integral  is understood as a sum of the integrals
over the smooth pieces of the boundary.
  The last term on the r.h.s. is due to the
 most external connected component of the
  boundary $\p \CD_{k}$ of the domain $\CD_{k}$
 having 2$n_{k}$ edges with angle $\pi /2$; their
contribution to the global geodesic  curvature is $\pi n_{k}$.

Now let us consider the third term $\CA '''$.
After integrating by parts the integrand
 in \hatn\ turns to $ dx (\xi) \bigwedge
d \varphi (\xi)$.
It is easy to see that  the contribution of each domain wall is
$ i {p_{0} \over 2\pi}
 (x_{{\rm right}} - x_{{\rm left}})\varphi _{\rm global}$ where
$\varphi_{{\rm global}} $ is the angle swept by the normal
vector $\hat n (\xi)$ along the domain wall
 (it is equal to the integral of the geodesic curvature).
Adding the contributions of all domain walls we find
\eqn\zzzzz{\eqalign{
 \CA ''' + i {p_{0} \over 2 \pi} \int _{\p \CM}
d\l (\xi) x(\xi) \hat K(\xi)  & = i {p_{0} \over 4 \pi }
\int _{\CM}d^{2}\xi \p _{a} x(\xi) \varepsilon  ^{ab} \varepsilon  ^{cd}
 \hat n _{c} \p _{b} \hat n _{d} \cr
 & =
 i {p_{0} \over 2\pi} \sum _{k}
 \int _{\p \CD _{k} } d\l (\xi) \hat K(\xi) x(\xi)\cr}}
Collecting the three terms
  \actn , \gb\  and \zzzzz\
 we arrive at the following action for given loop configuration
\eqn\actot{\CA = \CA ' +\CA '' +\CA ''' = K_{0} \l _{\rm tot} +
i {p_{0} }  \sum _{k } x_{k} [2 - c_{k}- {1 \over
2} n_{k}]}
Thus the Boltzmann weight $e^{\CA}$ of each domain wall configuration
depends only on its topology and the total length of the loops.
The action \actot\ can be simplified further by performing the sum over
the heigths $x_{k}$ of the domains $\CD_{k}$.
We have to
calculate the sum
\eqn\sumx{\Omega = \sum _{x_{k}}  e^{i  p_{0} \sum _{k} x_{k}
(2-c_{k}-{1 \over 2 } n_{k}) }}
The calculation is performed in the same way as in the case of the
closed string \Iade\ \Idis . We will exploit the fact that the system
of domain loops and open lines   on the disk has a tree-like structure.
Let us start with a domain $\CD _{k}$ on the top of the tree, i. e.,
a simply connected one.
It is represented by a vertex with a single line (tadpole) of the corresponding
graph.  Consider first the case when $n_{k}=0, c_{k}=1$  when
 the boundary is a closed loop. Then the sum over $x_{\rm inside}
=x _{k}$ yields
\eqn\sumxcd{\sum _{x_{ \rm inside}= x_{ \rm outside}\pm \pi}
\ \ \ \ e^{i p_{0} x_{ \rm inside}} \ \  = 2 \cos ( \pi p_{0}) e^{i
p_{0} x_{ \rm outside}}}
But $x_{ \rm outside}$ is the $x$ coordinate of the surrounding domain.
Therefore the result of the summation is a factor $2 \cos (\pi p_{0}) $
and a reduction  by one of the number of connected
boundaries ( $c \rightarrow c-1$)
of the surrounding domain.
Proceeding in the same way we can eliminate one by one all loops until we
arrive at a configuration  (a ``rainbow diagram'')
containing only open lines ending at the
boundary of the world surface. Each domain $\CD_{k}$ is characterized by the
number $n_{k}$ of the domain walls along its boundary ($c_{k}=1$).
 This configuration has again a structure of a
tree and we can sum over $x$ as before
starting with the  domains on the top of the tree, i.e., these
whose boundary is formed by a single line ($ c_{k}=1, n_{k}=1$).
The sum over the $x$ coordinate of such domain   yields
 a factor
$ 2 \cos ({1 \over 2}\pi p_{0}) $ and eliminates the term
associated with its boundary
\eqn\sososo{\sum _{x_{ \rm inside}= x_{ \rm outside}\pm \pi}
\ \ \ \ e^{ {1 \over 2}
i  p_{0} x_{ \rm inside}} \ \  = 2 \cos ({1 \over 2}\pi p_{0}) e^{
{1 \over 2}i p_{0} x_{ \rm outside}}}
After repeating this procedure several times  we eliminate all domain lines.
Thus the sum over the embeddings produces the following
 weight of each configuration of domain walls
\eqn\factr{\Omega =  \Big(2 \cos (\pi p_{0})\Big)^{\# \ {\rm loops}}
 \Big(2 \cos ({1 \over 2}\pi p_{0}) \Big) ^{\# {\rm open \  lines }}}
The sum over the last coordinate
 yields an infinite factor which is the volume of
the embedding space.

Summarizing, we arrived  at a modified loop gas model on the random surface.
Its partition function is a sum over configurations of nonintersecting
loops and open lines ending at the boundary
\eqn\partflg{\eqalign{
{\cal Z} =
& \sum_{\rm surfaces} \ \ \ \sum
_{\rm loop \ configurations} \cr
&   e^{-2K_{0}\l_{\rm tot}}   \Big( 2\cos(\pi p_{0}/2) \Big)^
{ \# {\rm open\ lines}}
\Big( 2\cos (\pi p_{0}) \Big) ^{\# {\rm loops}} \cr}}

The construction of the generalized loop gas can be  made more explicit
by  discretizing the measure over random surfaces as prescribed in \Iade .
The only difference is that the curvature is concentrated at the sites
of the lattice and the Gauss-Bonnet theorem  degenerates to the
Euler formula.

\newsec{Loop equations for the open string}

In order to exploit the definition  \partflg\  we have to give a meaning
of the functional integral over surfaces.
It is convenient to take the two sums in \partflg\
 in the opposite order: first
to  fix the topology of the configuration and the lengths of all lines,
 and then  perform the sum over the geometries of the
connected pieces of the surface (the ``windows'').  Each ``window''
contributes a factor depending only on the length of its boundary.
Finally we integrate over the lengths of the lines and sum over all
topologies. This sum can be most easily performed using equations of
Dyson-Schwinger
 type \Iade . Below we will use the continuum formulation of the
Dyson-Schwinger equations proposed in \ref\kaza{V. Kazakov, {\it
Mod. Phys. Lett} A 4 (1989)2125}.

In order to obtain a closed loop equation we have to consider a disk
with only one pair of Dirichlet and Neumann boundaries with lengths
$\l$ and $\tl$.
It seems that the only consistent way to avoid loops touching the
Neumann boundary is to have an open line end at each point.
If we are using a lattice regularization, this means that a line is
ending at the middle of each bond forming the Neumann boundary
(see Appendix A).

Let $V(\tl ,\l) $ be the partition function of the disk with such mixed
boundary conditions\foot{We denote this amplitude by $V$ saving the letter
$v$ for the corresponding renormalized amplitude}. It is related
to the functional  integral \fint\  with  $n=1$
by\foot{When $n=1$ the open string amplitude does not depend on the
position $x$ of its only Dirichlet boundary}
 \eqn\iaiaiss{V(\tl , \l )= \int _{0}^{\infty}dA e^{-\L A}
Z(A;\tl , \l , x)}
An infinitesimal deformation of the Neuman boundary
at its endpoint  (one of the points separating the two boundary conditions)
singles  out the line starting from this point which splits the world
surface into two pieces.
The loop equation follows from the geometrical decomposition of  the
disk shown in \fig\loooo{The geometry of the loop equation for the
Neumann-Dirichlet disk. The fat line represents the Neumann boundary.}
\eqn\leqn{{\p \over \p \tl} V(\tl ,\l ) = 2\cos(\pi p_{0}/2) \int _{0}^{\tl}
 d\tl '
\int_{0}^{\infty} d\l '  e^{-2K_{0}\l '} V(\tl ', \l ')\
 V(\tl -\tl ', \l +\l ')}
Eq. \leqn\ has a clear geometrical meaning. It sums up the rainbow diagrams
with an additional structure:  the space between its lines
is occupied by surfaces with loops.
Note that this loop equation determines only the dependence on $L$;
therefore it has to be complemented  with another equation
specifying the dynamics of closed loops. The missing information
can be supplied by fixing $W(\l)=V(0, \l)$
which is exactly the partition function of a disk with
Dirichlet boundary conditions.
It satisfies a loop equation \Iade\ - \Idis \ of the type
\eqn\Iiii{\eqalign{
-U'\Big({\p \over \p \l }\Big) W(\l)
=& \int _{0}^{\l} d\l ' W(\l ') W(\l -\l ' ) \cr
&+ 2\cos (\pi p_{0}) \int _{0}^{\infty}d\l  ' e^{-2K_{0}\l '} W(\l ')
W(\l +\l ') \cr}}
where $U'\Big({\p \over \p \l }\Big) $
is some local (differential) operator describing an infinitesimal
deformation of the boundary of the disk.

As usual, in order to turn the convolution in \leqn\
into a product, we introduce
the Laplace transform
\eqn\laps{\V(\tt ,t)= \int _{0}^{\infty} dL\int _{0}^{\infty} d\l
e^{-\tt \tl -t\l} V(\tl ,\l)}
and eq. \leqn\ turns to
\eqn\laplim{\tt \V(\tt ,t) -\W(t)= {\o \over 2\pi i} \oint {dt' \over t-t'}
\V(\tt ,t')\V(\tt , 2K_{0}-t')}
where $\tt$ plays the r\^ole of
 bare mass of the ``quarks'' at the ends of the open string
and
 \eqn\lloo{\W(t)= \int _{0}^{\infty} d\tt \V(\tt ,t)
= \int _{0}^{\infty} d\l \  e^{-t\l}W(\l)}
The contour integral in \laps\ goes around the singularities of $\W(\tt ,t)$
and leaves outsides the singularities of $\W(\tt , 2K_{0}-t)$.
Similarly, the Laplace transform of \Iiii\  reads
\eqn\liii{\W(t)^2={1 \over 2\pi i} \oint {dt' \over t-t'}
\W(t')^{2} [U'(t')-2\cos (\pi p_{0}) \W(2K_{0}-t')]}
All these loop equations can be derived in a rigorous way starting from
the lattice version of the model (see Appendix A).

The loop amplitude $V(\l, \tt)$ can be considered as the classical
background field in an open string field theory. It satisfies a
mean-field type equation which is equivalent to eq. \leqn .
This equation is derived by cutting the world sheet along the most internal
open lines as shown in \fig\altrp{The geometry of the classical field equation
for the open string tadpole}. In this way the  amplitude $V$ can be
expressed as an integral of the product of a $W$-amplitude and
a number of $V$ amplitudes\foot{By $V(\l , T)$ we denote the Laplace image
of \iaiaiss\ w.r. to $\tl$ ; it depends on $T$ through the factor
exp ($\#$ open lines)}
\eqn\tdcl{\eqalign{
& \V(t,\tt ) \cr
&= \sum _{n=0}^{\infty} \intoi d\l e^{-t\l}
\prod _{k=1}^{n} \Bigg( {d\l_{k} \over \tt} e^{ -2K_{0}\l_{k}}
\o V(\l_{k},\tt) \Bigg) \
W(\l +\l_{1}+...\l_{n}) \cr
&=\int _{M}^{\infty} {dt' \over \pi (t+t')}\
{\i \W(2K_{0}-t') \over 1- \o V( \tt , t')/\tt }\cr}}

It is known \Inonr\ \Idis\ \ref\mspaper{I. Kostov and M. Staudacher,
Multicritical phases of the O(n) model on a random lattice, Saclay
and Rutgers preprint SPhT/92/025, RU-92-6, March 1992}
that depending on the explicit form of the operator $U(\p /\p \l)$
one can achieve different critical regimes at the critical temperature
$K^{\ast}$ of the loop gas on the random surface.
Here we will consider in details the so-called  {\it dense phase}
corresponding to the simplest choice $U'(\p / \p \l)= \p / \p \l$.
In this phase the loops fill the world surface densely, without
leaving space between them. One of the peculiarities of the dense
phase is that the fractal dimension of the Dirichlet boundary
is larger than one: $1/\nu =1/(1-|p_{0}|) =1/g$.
The {\it dilute phase} of the loop gas corresponds to (multi)critical
potential $U$  \Idis\mspaper .  The potential is tuned so that the
area of the world surface not occupied by loops also diverges.
The equation for the loop amplitudes is the same for both phases but the
scaling of the cosmological constant is different. In the dilute phase
the fractal dimension of the Dirichlet boundary is $1/\nu = 1$.
In the  Coulomb gas picture the dense and dilute phases are related by
a duality transformation $g \to 1/g$.

\newsec{Solution of the loop equation in the scaling limit}

We will follow the method  worked out in \Iade\-\Idis\
for solving eq. \Iiii\ directly in the continuum limit, and
apply it to our master equation \laplim .

Let us first recall the solution of eq. \Iiii .
We expect that $W(t)$ has a cut $t_{L},t_{R}$ along the real axis of
the $t$-plane (on the first sheet of its Riemann surface). The contour of
integration in \liii\ goes around this cut.  If we replace in \liii\
$t$ with $2K_{0}-t$ the integrand will not change, but the contour of
integration will envelop the cut $[2K_{0}-t_{R}, 2K_{0}-t_{L}]$
of the function $W(2K_{0}-t)$. Therefore, adding these  two equations,
we integrate along both contours which form together a contour
surrounding all singularities of the integrand. Applying the Cauchy theorem we
find the following functional equation for $\W
(t)$ (we consider the simplest
 differential operator $ U'(t)=t$)
$$ \W(t)^{2}+\W(2K_{0}-t)^{2} + 2 \cos (\pi p_{0})
\W(t)\W(2K_{0}-t) $$
\eqn\wwt{ = t\W(t) +(2K_{0}-t)\W(2K_{0}-t)-2}
Taking the imaginary part of \wwt\ and knowing that ${\rm Im}W(t) \ne 0$
along the cut, we arrive at a linear Cauchy-Riemann problem:
$${\rm Re} \W(t) + \cos (\pi p_{0}) {\rm Im}\W(2K_{0}-t)=t/2,
 \ \ \ t \in [t_{L},t_{R}] $$
\eqn\cochrim{{\rm Im}W(t)=0,\ \ \ \ \ \ \ \ \ t \not\in [t_{L},t_{R}]}

If we take eq. \wwt\ at the symmetry point  $t=K_{0}$ we obtain
\eqn\Iv{ \W(K_{0}) = {2 \over K_{0} + \sqrt{K_{0} ^{2}
-4(1+\cos (\pi p_{0}))}}}
In the dense phase the temperature $2K_{0}$ of the loop gas is also the bare
cosmological constant since the total length of the loops is equal to the
area of the surface. The singularity of eq. \Iv\ gives its critical
value
\eqn\crrr{K_{0} \ \ \to \ \
K_{0}^{\ast} = 2\sqrt{1+ \cos {\pi p_{0}} } =2\sqrt{2}\sin
 (\pi g/2)}
At that point the two cuts touch each other:
\eqn\tuch{t_{R}^{\ast}=2K_{0}^{\ast} - t_{R}^{\ast} = K_{0}^{\ast}}

In order to explore the  vicinity of the critical point
we  blow up, as usual,  the infinitesimal vicinity
of the point $t=K_{0}^{\ast}$ by introducing a cutoff parameter
$a$ playing the role of elementary length along the boundary
\eqn\aaaa{
t = K_{0}^{\ast}+az,\
\ \ t_{R}= K_{\ast} -a M }
The parameter $z$  is coupled to the renormalized length of the boundary
(a boundary cosmological constant)
and $M$ is the renormalized position of the cut.
Note that the characteristic length of a loop grows near the critical
point as $(Ma)^{-1}$.

Since the singular part of the loop amplitude behaves for $M=0$ as
\ref\god{M. Gaudin and I. Kostov, \pl B220(1989) 200}
\eqn\godd{\W(t) \sim (az)^{g}, \ \ \ g=1-|p_{0}|}
we define the scaling part of $\W$ as
\eqn\scprt{\W(t)=W^{\ast} +  A_{g} a^{g} \w(z)}
where $\W^{\ast}$ is the critical value  of $\W$ at $t=K_{0}^{\ast},
K_{0}=K_{0}^{\ast} $  and $A_{g}$ is a constant factor depending on the
normalization of $\w$.

Finally, we introduce the renormalized cosmological constant
\eqn\cosm{
 K_{0} =K_{0}^{\ast}
+ B_{g} a^{2\nu} \L}
where $B_{g}$ is an appropriate
 constant and  $\nu$ has the meaning of the inverse fractal dimension of
the Dirichlet boundary, if the dimension of the world sheet is assumed
to be 2.
Since $\L$ is the only parameter in the theory,
 we expect that $ M^{2\nu} \sim \L $.
To determine $\nu$ we note that from \aaaa -\scprt\
(for $1/2<g<1$)
\foot{In this interval $ K_{0} =K_{0}^{\ast}$ up to terms of
higher than linear order in the cutoff $a$}
\eqn\nyuu{\eqalign{
\W(K_{0})&=\W^{\ast} +  A_{g}  a^{g}  \w(0) \cr
                  & =\W^{\ast} -  (2 /K_{0}^{\ast})^{3/2} \sqrt{ B_{g}\L} \cr}}
and, sinse $\w(0) \sim M^{g}$, we find $\nu = g$.

The renormalized loop amplitude $\w(z)$ has a cut $-\infty <z<-M$
and satisfies the following equation which is a direct consequence of
\cochrim
\eqn\wfw{
\eqalign{
 {\rm Re} \w(z) - \cos (\pi g) \w(-z) & =0, \ \ \ z \le -M \cr
{\rm Im}\w(z) &=0, \ \ \ z \ge -M \cr}}
If we parametrize $z$  by means of a new variable $\tau$
\eqn\zzz{z = M\cosh \tau  }
the reflection $z \to -z \pm i0$  corresponds to
$\tau \to \tau \pm i \pi$ and \wfw\ is replaced by
\eqn\wtw{ [e^{i \pi \p / \p \tau}+ e^{-i \pi \p / \p \tau} -
2\cos (\pi g) ] \w(z)=0,\ \ \ g=1-|p_{0}|}
with an evident solution \foot{This normalization
corresponds to  $A_{g}=2^{3g/2-1/2}(1-g)^{g-1}[\sin (\pi g/2)]^{-g-1}$}
\eqn\wwzzw{\eqalign{
\w(z) &= -M^{g} {\cosh (g\tau) \over \cos (g\pi /2)} \cr
& =-{(z+\sqrt{z^{2}-M^{2}})^{g}+(z+\sqrt{z^{2}-M^{2}})^{g} \over
2\cos (g\pi /2)}
\cr }}
We have normalized the solution so to have
\eqn\ser{\w(0)=-M^{g}}
Then  by \nyuu , with $B_{g}=[K_{0}^{\ast}/2]^{3} A_{g} ^{2}$ , the
 relation between $M$ and $\L$  is just
\eqn\mmsss{\L = M^{2g}, \ \ \ \ \ \  0<g<1}
The function \wwzzw\ has a cut $[-\infty, -M]$ on its physical sheet
whereas the cut $[M, \infty]$ appears only on the next sheets.

In the same way we can analize the dilute phase of the loop gas $(g>1)$.
We would obtain the same expression \wwzzw\ for the loop amplitude but
the scaling of the cosmological constant will be different:
$\L = M^{2}$ \Inonr\Idis\mspaper .  The scaling of $\L$ in both phases of
the loop gas is determined by the dimension $d_{D}=1/\nu$ of the Dirichlet
boundary, with $\nu$ given by \nuuu
\eqn\sclg{\L= M^{2\nu}, \nu = \cases{
g, \ \ \ \ \  &if $g<1$ \cr
1, \ \ \ \ \  &if $g>1$ \cr}}

Let us now consider the loop equation  \laplim\ for the open string.
We choose to work in the dense phase, but all calculations can be easily
extended to both phases of the loop gas.
Again, after symmetrization
 w.r.t. the reflection $t \to 2K_{0}-t$
\eqn\Iijk{\eqalign{
&T [\V(T,t) + \V(T, 2K_{0}-t)] =\W(t) + \W(2K_{0}-t)\cr
&+  \o \V(T, t) \V(T, 2K_{0}-t)\cr}}
We have assumed that $\V(t)$ has the same cut as $\W(t)$.

In order to determine the critical value of $T$ we consider eq.
\Iijk\ at the point $t=K_{0}$  where it becomes algebraic
\eqn\Iivv{2T\V(T, K_{0}) = 2\W(K_{0}) + \o \V^{2}(T, K_{0})}
Using \nyuu\
and  dropping  all powers of the cutoff $a$ higher than $a^{g/2}$
we write its solution in the vicinity of the critical point as
\eqn\solu{\eqalign{
\V(T,K_{0})&= {T-\sqrt{T^{2}-2\W(K_{0}) \o} \over \o} \cr
& \approx \V^{\ast}-
  a^{g/2} \xo \sqrt{ A_{g}( \mu + 2\sqrt{\L})}}}
where
\eqn\lala{\V^{\ast} = T_{\ast} =\sqrt{{2W^{\ast}
\over \o}}}
is the critical value of the open string amplitude and
\eqn\soluu{ 2 a^{g} \mu =  A_{g} [ T^{2}-T_{\ast}^{2}]}
 is the renormalized mass at the endpoints
of the open string (the parameter coupled to the length of the Dirichlet
boundary).

As before, we retain in the scaling limit only the  singular part of
$\W(t)$:
\eqn\sptw{\V(T,t)=\V^{\ast} + \ a^{\alpha} \sqrt{{A_{g} \over \o}} \v(\mu ,z)}
Comparing \sptw\ and  \solu\ we find
\eqn\votkak{\v(\mu , 0)= - \sqrt{2(\mu + \sqrt{\L})}}
Then, throwing away the higher order terms we obtain from \Iijk\
$\alpha = g/2$ and
\eqn\contne{
\w(z) +\w(-z) + \v(\mu ,  z) \v(\mu , - z) =2\mu }
At $z=0$ this equation reproduces \votkak .

This equation is compatible with  the integral equation \tdcl\ in the
scaling limit. Indeed,
introducing
the scaling variables according to \aaaa\ \scprt\ and \sptw\ we find
the integral equation
\eqn\tddpl{\v(z, \mu)= \int _{M}^{\infty}   { dz_{1} \over \pi
(z+z_{1})} {\rm Im}
{w(-z_{1}) \over v( z_{1}, \mu)}}
 Taking the imaginary part of \tddpl\
along the cut we find
\eqn\eqqlv{\i \v(z,\mu )= - {\i \w(z) \over \v(z, \mu )} , \ \ \ \ \  z <-M}
which gives the imaginary part of \contne\ .

Exactly the same equation can be obtained for the dilute phase of the loop
gas on the world sheet  which corresponds to the choice $g>1$. The only
difference is that
the cosmological constant is replaced by $\L = M^{2}$.
All further arguments are valid for both regimes.

If we parametrize $z$ by \zzz\ and $\mu$ by
\eqn\idinah{ \mu =  M^{g} \cosh(g \s); \ \ \ \ \  \v (\mu , z)= v(\s , \tau)}
 eq. \contne\  becomes
\eqn\blyad{v( \s , \tau +i \pi) v(\s , \tau )
=M^{g}\Bigg( { \cosh [g(\tau +i\pi)] + \cosh (g\tau)
 \over \cos (g\pi /2)} + 2\cos (g\s ) \Bigg)}
After shifting  $\tau$ to $\tau +i\pi /2$ , eq. \blyad\ becomes
\eqn\pizda{v(\tau + i \pi /2) v(\tau -i \pi /2) =
4  M^{g} \cosh [{g \over 2}(\tau + \s) ] \cosh
[ {g \over 2} (\tau - \s)]}

In the limit $\L=0,\mu =0$ the solution of \pizda\ is
\eqn\crit{\v(z)= (2z)^{g/2}}

For  for nonzero $\L$ and $\mu $
but for some particular values of $\s$
\eqn\suka{\s= \pm i \pi /2 , \pm i\pi /2 \pm i \pi /g}
eq. \pizda\ has a solution in elementary functions
\eqn\nadoelo{\eqalign{
&v( \pm i \pi /2 , \tau) = - 2 M^{g/2} \cosh (g\tau /2 ) \cr
& v( \pm i\pi /2  \pm  i \pi /g  , \tau) = -
 2 M^{g/2} \sinh (g\tau /2 ) \cr }}

In order to solve eq. \idinah\ in the general case let us take the
logarithm of both sides to obtain a linear equation on
\eqn\taau{u(\s ,\tau)= \log v(\s ,\tau) }
of the form
\eqn\ttau{ (e^{i {\pi \over 2} {\p \over \p \tau}}
+e^{ -i {\pi \over 2} {\p \over \p \tau}}) u(\s ,\tau) =
\log [2\cosh ( g(\tau +\s)/2)] + \log
[2\cosh ( g(\tau -\s)/2)] }
It is easy to solve it by performing a Fourier transform
which gives an integral representation for $u(\s ,\tau)$
\eqn\au{ u(\s ,\tau)= u(\tau ,\s)= f(\tau + \s ) +f(\tau - \s)}
\eqn\auu{f(\tau)= f(-\tau)= - {1
 \over 4} \int _{-\infty}^{\infty} {d\omega \over \omega}
{ e^{i\omega \tau} \over
 \cosh (\pi \omega /2) \sinh (\pi \omega /g)}  }
The  ambiguity due to the singularity at $\omega =0$
is lifted by imposing the condition $f( \tau ) \to g |\tau|/4 $
when $\tau \to \infty $.
By deforming the contour of integration and applying the Cauchy theorem
we can write the integral \auu\ as the following formal series
which makes sense for ${{\rm Re}}\tau$ positive
\eqn\uau{f(\tau)= {g \over 4}\tau  +{1 \over 2}
 \sum _{n=1}^{\infty}{(-1)^{n-1}
\over n} {e^{-gn\tau}
\over \cos (\pi g n/2)} + \sum _{k=1} ^{\infty}{(-1)^{k-1} \over
2k-1} {e^{-(2k-1)\tau} \over  \sin [(2k-1)\pi /g]}}

When $g$ is rational, this series for $df/d\tau$ can be easily summed up.

Consider for example the case when  $g=p/q$ with
$p,q$ co-primes and $p$ even. In this case $q$ is automatically odd.

Representing the summation indices in \uau\ as
\eqn\sspq{\eqalign{
n&=2qN+\bar n, \ \ N=0,1,2,...; \bar n = 1,2,...,2q \cr
k&=pK+\bar k, \ \ \ K=0,1,2,...; \bar k = 1,2,...,p \cr}}
we arrive at the following expression
\eqn\Ivan{\eqalign{
{df(\tau) \over d\tau}& =
{1 \over 4 \sinh (p \tau)} \Bigg( g\cosh (p\tau)
 + g \sum _{\bar n =1}^{2q-1}(-)^{\bar n-p/2}
{\cosh [g(q-\bar n)\tau] \over \cos [g(q-\bar n)\pi /2]} \cr
&+2 \sum _{\bar k =1}^{p} (-)^{\bar k}
{\cosh[(p+1-2\bar k)\tau] \over \sin [(p+1-2\bar k)\pi /g]} \Bigg)\cr}}
In the simplest case $g=2 \ \ (p=2, q=1)$ this expression reproduces
the result
\eqn\gdve{f(\tau)=\log (\cosh {\tau \over 2});
\ \ \ \ \ \  {df(\tau) \over d\tau} = {1 \over 2 } { \rm th} (\tau /2)}
which can be easily obtained directly from the integral \auu .

If $g=p/q$ with $p$ odd, some of the coefficients in
 this series become infinite.
This happens when $ng=2k-1$. It is easy to see that the divergent
coefficients appear in pairs and the contribution of each such pair is
finite:
\eqn\pr{{(-1)^{n-1} \over n} {e^{-gn\tau}
\over 2\cos (\pi g n/2)} +{(-1)^{k-1} \over
2k-1} {e^{-(2k-1)\tau} \over  \sin [(2k-1)\pi /g]} \to
{(-1)^{n+k} \over n} {\tau \over \pi}e^{-(2k-1)\tau} }
The r.h.s. represents the limit of the l.h.s. when $g \to (2k-1)/n$.
For example, for $g=1$ we obtain:
\eqn\auau{f(\tau)= {1 \over 4} \log (\cosh \tau) +{\tau \over \pi}
{\rm arctg} e^{-\tau } -{1 \over 2 \pi} \sum_{0}^{\infty} {e^{-(2n+1)\tau}
\over
(2 n +1)^{2}}}
This result is already inexpressible in terms of elementary functions,
unlike the formula for the derivative $df/d\tau$.

Let us note that our disk amplitude
 $v(\sigma,\tau)$ being represented in the integral form \auu\
 is remarkably similar to the $S$-matrix of the $O(n)$-vector model
with $n=-2\cos  (\pi g)$
on the regular lattice presented in the paper
 \ref\smir{F.Smirnov, Santa Barbara  preprint NSF-ITP-90-237 }.
\foot{notice a misprint there:the factor $1/k$ was missing there in
the  integral}.Our $\tau$-parameter corresponds to the rapidity parameter in
the two-particle $S$-matrix. Eq. \pizda\ is analogous to the unitarity
condition on the $S$-matrix.
 This S-matrix was first calculated in
 \ref\zam{A.Zamolodchikov, Mod.Phys.Lett. A6 (1991) 1807}
 in terms of an infinite product of gamma functions, which
we can use for our amplitude as well. Expanding $\cosh{\pi \omega \over 2}$ in
\auu\ in the exponents and performing the integration we obtain:
\eqn\gam{v(\tau,0)=e^{2 f(\tau)} =
e^{g \tau \over 2} \prod_{k=0}^{\infty}
{\Gamma({1 \over 2} + g{k+3 \over 4}  +{\tau g \over 2 \pi i})
\Gamma({1 \over 2} + g{k+3 \over 4}  -{\tau g \over 2 \pi i}) \over
\Gamma({1 \over 2} + g{k+1 \over 4 } +{\tau g \over 2 \pi i})
\Gamma({1 \over 2} + g{k+1 \over 4 } -{\tau g \over 2 \pi i})}
{\Gamma^2({1 \over 2} + g{k+1 \over 4 })  \over
\Gamma^2({1 \over 2} + g{k+3 \over 4 })}}
The exponential factor in front of the product depends on how we treat the
singularity at $\omega=0$ in this integral. It is defined through the
asymptotics \crit\ of $v({\tau})$.

     It is not clear whether this coincidence is accidential or it reflects
some
deep relationship between the $O( n)$-vector
 field in the flat and fluctuating metric
of two-dimensional space, respectivly.
 May be the representation of the model in the flat space in terms
of the effective Sine-Gordon theory presented in \smir\ can shed some light
on this strange fact.

\newsec{Boundary operators}

It is very convenient to regard the loop amplitudes as expectation values of
operators creating boundaries on the world sheet.
  For this purpose we are going  to introduce the following notations.
Denote by $\CO (\l)$  the operator creating a closed Dirichlet
 boundary of length $\l$ on the world sheet.
 The expectation value of this operator is nothing but
 loop amplitude \wwzzw\
\eqn\wzwzw{\w(z)= -{\p \over \p z}
\langle \CO (z) \rangle, \ \ \CO(z)=\int _{0}^{\infty}
d\l \ e^{-z\l} \CO (\l)}
(The derivative comes from the loop amplitude being defined with a
marked point on it.)

Similarly,  by $ \tilde  \CO (\tilde \l )$  we denote the operator
creating closed Neumann boundary of length $\tilde \l$. Its expectation
value is  the loop
amplitude with Neuman boundary condition
\eqn\nene{\tilde w(\mu)= -{\p \over \p \mu} \langle \tilde \CO(\mu)\rangle ,
\ \ \   \tilde \CO(\mu )= \int _{0}^{\infty} d \tilde \l
 \ e^{-\tilde \l \mu}  \tilde \CO(\tilde \l) }

Once
a Dirichlet boundary exists, one  can define a boundary  operator
 $\t \CC(\tl)$ creating an open  Neumann boundary
 of length $\tl$ at some point.
 In a similar manner we define the operator
 $\CC (\l)$ creating open Dirichlet boundary of length $\l$
 at some point on the
Neuman boundary.
By construction
\eqn\coco{
 \tilde  \CC (\tilde \l ) \CO(\l)=
\CC(\l)  \tilde  \CO (\tilde \l ) }
The disk amplitude $\v(z,\mu)$ with Dirichlet-Neumann boundary
conditions  is the expectation value of any of the products \coco
\eqn\ozozo{\v(z, \mu)= \langle \tilde \CC(\mu) \CO(z) \rangle
 =\langle \tilde \CC(\mu) \CO(z) \rangle}
Here we used the notations
\eqn\lapla{
   \CC(z)= \int _{0}^{\infty}
d  \l \ e^{- z  \l} \CC (\tilde \l),
\ \   \tilde \CC(\mu)= \int _{0}^{\infty}
d \tilde \l \ e^{- \mu \tilde \l} \tilde \CC (\tilde \l)}

The operator $\CC(\l)$ (resp.  $\tilde \CC(\tilde \l)$ )
 creating open
Dirichlet (resp. Neumann)
boundary   can  be expanded as  an infinite series of local boundary operators
in the same way as the  loop operator in the closed string is
expanded as a series of operators creating microscopic loops
\ref\mss {G. Moore, N. Seiberg and M. Staudacher, \np  \ 362(1991)665}.
(The boundary operators in the framework of the Liouville theory have been
studied in \Mar . )

Consider first the limit of large $\mu$ corresponding to small Neumann
boundary and
 expand  $\v (\mu, z )$ in negative powers of
$\mu$
\eqn\expn{\eqalign{
v(\tau, \s)& =M^{g/2} e^{g\s /2} \ [1+ {\cosh (g\tau) \over
\cos (g\pi /2)} e^{-g\s} + {\cosh \tau \over \sin (\pi /g)} e^{-\s} + ...] \cr
&= \sum _{k,n=0}^{\infty} \t C_{k,m}(\L , z ) \mu^{{1 \over 2} -k-n/g}
\cr}}
with
\eqn\coef{\tilde C_{0,0}=1,
\ \  \tilde C_{1,0}={M^{g}\cosh (g\tau ) \over \cos (\pi g/2)}= \w(z),
\ \ \tilde C_{0,1}= {M\cosh \tau \over \sin (\pi /g)} \sim z, ...}
The coefficients in \expn\ can be interpreted as expectation values of local
boundary operators \foot{The operators $\CC_{0,0},\CC_{0,1}$,... are
not, strictly speaking, local operators. They are ``boundary operators''
for the Dirichlet boundary }
\eqn\bop{\t C_{k,n}(\L , z)= \langle \t \CC_{k,n}  \CO (z) \rangle}
so that \expn\ would imply the following expansion of the operator
$\t \CC(\mu)$ creating Neuman boundary as an infinite series of
local scaling operators
\eqn\llll{\t \CC(\mu)= \sum _{k,n} \t
\CC_{k,n}\mu ^{{1 \over 2 }-k -{n \over g}}}
The leading nontrivial coefficient in \expn\   is just the amplitude
$\w (z)$  of the closed string and the corresponding boundary operator is
\eqn\ddz{\t \CC_{1,0}=-{\p \over \p z}}

Let us define the dimension  $ \delta$ of a local
boundary operator $\tilde \CC$
acting at a point of the Dirichlet boundary.
The mean value  $  \langle \tilde \CC \CO(z) \rangle = F(\L,z,\mu)$ of such
 an operator is assumed to have the following scaling property
\eqn\scccll{ F(\rho ^{2}\L , \rho ^{d_{D}} z, \rho ^{d_{N}} \mu )
=\rho ^{ \alpha} F(\L , z, \mu)}
where
\eqn\scc{\alpha=(2-\gst) \chi -d_{D}(1- \t \delta) }

Note that we measure the dimension of the operator $\CC$ in units of
dimension of the Dirichlet boundary and not the world sheet.
Otherwise we would have an additional factor of $d_{D}/2$.
Let us  recall that the dimensions of the Dirichlet boundary
is $d_{D}=1/\nu$ and the dimension of the Neumann boundary is
$d_{N}= 1 /\tilde \nu = g/\nu$ if the dimension of the world sheet is 2.
The term $d_{D}$ (resp. $d_{N}$ ) comes from the fact that marking a point
on the boundary  breaks the cyclic symmetry and produces a factor of
length.
In the case of the topology of a disk (Euler characteristic
 $(\chi =1)$ ),  eq. \gstring\ yields
$\alpha ={g+ \t \delta \over \nu}$ and the dimension of the operator $\t \CC$
is related to $\alpha$ by
$ \t \delta = \alpha - g/\nu$.

Now let us examine the mean values $\t C_{k,n}$. It is easy to check that
they satisfy the scaling \scccll\ with   $\alpha =
(kg+n)/\nu$. Therefore the (boundary) dimensions of the corresponding local
operators are
\eqn\dddmms{\delta _{k,n}=(k-1)g+n}
The dimension of the operator $\t \CC_{1,0}$ is zero, as expected.

Let us consider the opposite limit $z \to \infty$ corresponding to small
Dirichlet boundary.
The corresponding expansion of $\v$ is obtained from \au\ and \uau\ with
$\s >\tau >0$
 From \au\ and \uau\  , assuming that  $\tau >\s >0$, we find
\eqn\expn{\eqalign{
v(\tau, \s)& =M^{g/2} e^{g\tau /2} [1+ {\cosh (g\s) \over
\cos (g\pi /2)} e^{-g\tau} + {\cosh \s \over \sin (\pi /g)} e^{-\tau} + ...]\cr
&
= \sum _{k,n=0}^{\infty} C_{k,m}(\L , \mu ) z^{({1 \over 2} -k)g-n}\cr}}
with
\eqn\coef{C_{0,0}=1,\ \
 C_{1,0}={M^{g}\cosh (g\s ) \over cos (\pi g/2)}\sim \mu ,
\ \ C_{0,1}= {M\cosh \s \over \sin (\pi /g)} = \tilde w(\mu), ...}

Analogously to the previous case we
   define the dimension $\delta$ of a local
 operator at some point on the Neumann boundary
 by the scaling properties of
 $  \langle \CC \tilde \CO(\mu) \rangle = F(\L,z,\mu)$.
In this case the power $\alpha$ in \scccll\  is related to the dimension
of $\CC$  by
\eqn\pwr{ \eqalign{
 \alpha =& (2-\gst)\chi -d_{N}(1-\delta) \cr
=&(1+g\delta)/\nu \cr}}
We  consider  the coefficients $ C_{k,n}$
as mean values of microscopic operators ${\CO} _{k,n}$ at
the Neumann  boundary which implies the expansion
\eqn\zubb{\CC(z)=\sum _{k,n} \CC_{k,n}\ z^{g({1 \over 2}-k) -n}}

The  coefficient $ C_{k,n}$ obeys the scaling \scccll\ with
$\alpha = (gk+n)/\nu $ and we find by \pwr\
\eqn\dmsnss{ \delta _{k,n} = k + {n-1 \over g}}
The identity operator is $\tilde {\CO}_{1,0} = - \p / \p z$.
Its expectation value $ C_{0,1}$  gives the loop amplitude
with Neumann boundary condition
\eqn\nloop{\tw (\mu) =C_{0,1}= {[\mu +\sqrt{\mu ^{2} -M^{2g}}]^{1/g}
+  [\mu -\sqrt{\mu ^{2} -M^{2g}}]^{1/g} \over 2\sin (\pi /g)}
}

Let us make the following remark.  The duality transformation of the
functional integral
with  the  gaussian  action  \pact\  leads to a similar action but
with $g$ replaced by $1/g$ and the Dirichlet and Neumann boundary conditions
exchanged. The symmetry of the function $u(\s, \tau)$ is a manifestation
of this duality symmetry. In this sense the dense $(g<1)$ and the diluted
($g>1$) phases of the loop gas are dual to each other.
In the dense phase the Neumann boundary has the classical
 dimension $d_{N}=g/ \nu =1$
 while the Dirichlet boundary has anomalous dimension $d_{D}=1/\nu = 1/g >1$.
In the dilute phase the Dirichlet boundary has classical
 dimension $d_{D}=1/\nu =1$
while the Neumann boundary has anomalous dimension $d_{N}=g/\nu = g >1$.
The loop amplitude with Neumann boundary condition is related  to that with
Dirichlet boundary condition by $z \leftrightarrow \mu , g \leftrightarrow
1/g$. Therefore the quasiclassical treatment (see, for example
\ref\seib{N.Seiberg, {\it Prog. Theor. Phys. Suppl.} 102 (1990) 319})
is applicable for the Neumann boundary when $g<1$ and for the Dirichlet
boundary when $g>1$ but not for both in the same time.

The operators involved in the expansions \llll\ and \zubb\ are not the
only boundary operators presented in the theory.
Each kind of boundary allows its special boundary operators.

Consider first the Dirichlet boundary. Since all points have the same
$x$-coordinate, it is kinematically impossible to introduce an order
operator $\exp (ipx)$. However, we can define a {\it disorder operator}
 \ $\t \chi _{[m]}$
with magnetic charge $m$ representing a discontinuity $\Delta x= m\pi$
at some point of the Dirichlet boundary.
Geometrically this operator is represented by a source of $m$ domain lines
starting at the same point at the boundary.
The expectation value of such an operator can be calculated by
decomposing the world sheet along the $m$ lines (\fig\grrrr{The
geometrical description of a disorder operator and the
corresponding decomposition of the world surface}):
\eqn\desoor{\eqalign{
&\langle \t \chi_{[m]} \CC(z) \CO(\mu)  \rangle \cr
&= \intoi d\l _{0}\intoi d\l_{1}...\intoi d\l _{m+1} \cr
& e^{-(\l_{0}+\l_{m+1})z}
v(\l_{0}+\l_{1},\mu) v(\l_{1}+\l_{2},\mu) ... v(\l_{m}+\l_{m+1},\mu)\cr
= &\int _{M}^{\infty}{ dz_{1}\over \pi}...\int _{M}^{\infty}{dz_{m+1}\over \pi}
{\i [\v(-z_{1},\mu)] \i [ \v (-z_{2},\mu)] ...  \i [ \v (-z_{m+1},\mu)]
\over (z+z_{1})(z_{1}+z_{2})...(z_{m}+z_{m+1})(z_{m+1}+z)}\cr}}
Since $\v(z,\mu) \sim z^{g/2}$, the whole integral scales as $z^{(m+1)g/2
-1}$. On the other hand, the amplitude with a marked point on the Dirichlet
boundary scales as $\p \v(z,\mu)/\p z \sim z^{g/2-1}$.
Comparing the two powers we find the dimension of the desorder operator on
the boundary
\eqn\bddo{\t \delta _{[m]}= mg/2 }

With the Neumann boundary the things stay in the opposite way.
The disorder operators do not make sense because there is already a
discontinuity at each point of the boundary.  However, the order
operator $\CV_{(p)}$ introducing a factor
 $\exp [i(p-{1 \over 2}p_{0})x]$  at some point
$\xi$ of the boundary can be defined perfectly well. Going to the
Fourier space and distributing the exponential factor among
the domain lines crossing the way between the point $\xi$ and the Dirichlet
boundary, we arrive at the following geometrical description of the
order operator with electric charge (momentum) $p$. The expectation value
 \eqn\ampp{\v_{(p)}(z,\mu)=\langle \CV_{(p)}\CC(\mu)\CO(z)
 \rangle }
 is equal to the statistical sum for the mixed
Dirichlet-Neumann loop amplitude with the fugacity of some of the
domain lines modified. Namely, all domain lines surrounding the point $\xi$
are taken with a factor $\cos (\pi p)/\cos({1 \over 2}\pi p_{0})$.
(We remind that the factor of $
 \cos({1 \over 2}\pi p_{0})$ has been absorbed in T)

The loop amplitude
\ampp\
satisfies the following integral equation
\eqn\tdpl{\v_{(p)}(z,\mu)= \int _{M}^{\infty} {\cos (p)
\over \cos ({1 \over 2}\pi p_{0})}  { dz' \over \pi
(z+z')}
{ \v_{(p)}(z,\mu)  {\rm Im}[\w(-z)]  \over [\v( z,\mu)]^{2}}}
which can derived in the same way as eq. \tdcl .
Eq. \tdpl\ can be considered as  the dispersion integral for an analytic
function with a cut $M<z<\infty$. Therefore along the cut we have
\eqn\eeee{
[\i \v_{(p)}(z,\mu) ]= - {\cos (\pi p)
\over \cos ({1 \over 2}\pi p_{0})}
{ {\rm Im}[
\w(z)]  \v_{(p)}(-z,\mu) \over [\v( -z,\mu)]^{2}} , \ \ \ z<M}
Inserting the relation
\eqn\tttt{{\rm Im}[\v(z,\mu)] + { \i [\w(z)]\over \v(-z,\mu)}=0}
 in \eeee\ we find
\eqn\locp{
\i [\v_{(p)}(z,\mu) ]= - {\cos (\pi p)
\over \cos ({1 \over 2}\pi  p_{0})}
{ {\rm Im} [ \v(z,\mu )]  \over \v( -z,\mu)}  \v_{(p)}(-z,\mu), \ \ \ z<- M }

For $z$  large $v(z,\mu) \sim z^{g/2}$ and therefore
 \eqn\aszg{    { \i [\v(z, \mu )] \over  \v(-z, \mu) } \to
\sin ({1 \over 2}\pi g)      = \cos  ({1 \over 2}\pi p_{0}), \
\ \ z \to -\infty}
Therefore at large $z$ the amplitude \ampp\ satisfies the functional equation
\eqn\jjja{ \i [ \v_{(p)}(z,\mu)]= \cos (\pi p) \v_{(p)}(-z,\mu)}
whose solution is any power $z^{\alpha}$ with $\alpha = \pm ( p -{1 \over 2})
+ {\rm even\ integer}$.
The leading power at $z \to \infty $ large can be fixed by the
requirement that when the momentum $p$ coincides with the background
momentum ${1 \over 2} p_{0}$, the amplitude
\ampp\  coincides with the expectation value of the
identity operator $-\p / \p \mu$
\eqn\pnula{\v_{({1 \over 2}p_{0})}(z,\mu)=- { \p \over \p \mu}
\v(z, \mu)}
 Therefore  $ \v_{(p)}(z, \mu)$ behaves for $z$ large  as
\eqn\asimzl{ \v_{(p)}(-z, \mu) \sim z^{ |p -{1 \over 2} p_{0}| -{g \over 2}}}

Comparing this with the asymptotics $\p \v /\p \mu \sim z^{g/2-g}$
we find
\eqn\dimor{\delta_{(p)}= { |p- p_{0}/2| \over g}}

 For finite $z$ \ampp\ is given by the infinite series
\eqn\ryad{\v_{(p)}(z,\mu) = \sum _{k,n}C_{k,n}^{(p)}(\mu , \L)
 z^{|p-{1 \over 2}  p_{0}|-g({1 \over 2}+k) -n }}

\newsec{Open string propagator and the spectrum of momenta}

All  amplitudes involving closed and open strings are defined
by imposing appropriate Dirichlet and Neumann boundary conditions
on the boundaries of a world sheet with given  topology.

 Any string amplitude can be decomposed into elementary pieces
(propagators and vertices) following the  logic of refs. \Imult\ and \Idis .

In this paper we concentrate ourselfs on the calculation of the open string
propagator. It will be obtained following the same steps
as in the case of the closed string propagator \Imult .
Contrary to our expectations, the case of open strings turned out
to be technically much more difficult than the case of closed strings.
We have found the spectrum of the propagator
but we   were not able to  obtain the  explicit form of the eigenstates.

   Before  considering the open string, let us repeat the major steps
of the calculation  of the closed string propagator, using the SOS model.
    One has to calculate a string-string amplitude with the world sheet
configuration of the loops as shown in \fig\ffrr{ Loop configuration
 for the closed string propagator}, with
non-contractable domain walls going around the cylindric surface.  In
this way we take into account the possibility for the closed string to
propagate in the $x$-space.

 Let $x$ and $x'$ be the coordinates of the two (Dirichlet) boundaries
of the cylinder. If we pass to the momentum space
the factor $e^{ i \pi p(x-x')}$ can be written as a product of factors
$e^{\pm i \pi p} $ associated with the domain walls wrapping the cylinder.
Taking into account the two different orientations, each such domain wall
acquires a factor $2\cos (\pi p)$. Further, the amplitude of each
elementary cylinder between two subsequent noncontractable
 domain walls
is \Iade \Idis
\eqn\eelcy{G_{0}(\l , \l ')= \sqrt{\l} {e^{- M(\l + \l ')} \over \l + \l '}
\sqrt{\l '}}
This amplitude describes  the deformation of the closed string from the
"in" state of  length $\l$ to the "out" state with a length $\l '$,
without a change of the $x$-space position.
     The whole propagator $G(p;\l,\l ')$ in the momentum space
can be  obtained by sewing
such elementary cylinders:
\eqn\prpgt{G(p;\l,\l ')=\sum _{n=0}^{\infty}
\int _{0}^{\infty} ...
\int _{0}^{\infty} {d\l _{1} \over \l _{1}} ...  {d\l _{n} \over \l _{n}}
[2\cos (\pi p)]^{n}
G_{0}(\l, \l _{1}) G_{0}(\l _{1}, \l _{2}) ... G_{0}(\l _{n}, \l ')}
     To calculate it we have to diagonalize $G_{0}(\l,\l ')$ in the
$\l$-space. This was done in
\mss\
 \eqn\diag{G_{0}(\l , \l ')=
 =  \int _{0}^{\infty} dE
\langle \l |E \rangle {1 \over 2\cos (\pi E)} \langle E| \l ' \rangle }
where
\eqn\kfun{\langle \l | E \rangle = {2 \over \pi} \sqrt{\pi E \sinh (\pi E)}
\ K_{iE}(M\l)  \ \ \approx   ( \l M)  ^{i E}, \l
 \to 0}
form a complete set of delta-function normalized eigenstates
\eqn\deltno{\int _{0}^{\infty} {d\l \over \l}
\langle \l |E \rangle  \langle E'| \l  \rangle
= 2 \pi \delta (E,E')}
It is convenient to introduce the Liouville variable $\phi = \log \l$;
then the
integration measure becomes uniform:
\eqn\meas{ {d\l \over \l}= d\phi ; \ \  \l = e^{\phi}}
The wave functions behave as plane waves in the limit  $ \l \to 0
\ \ (\phi = \to - \infty)$  and decay rapidly  when $\phi \sim \log (1 /M)$.
Therefore the $\delta -$function is produced only by the small-$\l$
behavior and the normalization coefficient is not affected by the form of
the eigenstates for $\l \sim 1/M$. This important feature of the
half-space quantum mechanics was emphasised and well explained in \mss .
It can help  to calculate directly the spectrum of the kernel $G_{0}$
 from its small-$\l$ behaviour.
Indeed, let us expand the r.h.s.  of \eelcy\ in $\l /\l '$ ,assuming
that $\l < \l '$.  One finds,  writing the series as the result of a contour
integration
\eqn\contr{G_{0} (\l, \l ') = \sum _{0}^{\infty} (-)^{n} \Big( {\l \over \l '}
\Big)^{n+1/2} = \int _{-\infty}^{\infty} {dE  \over 2\cosh (\pi E)}
 \Big({ \l \over \l '}\Big)^{iE}}

Once the irreducible part $G_{0}$ is diagonalized,
  the r.h.s. of \prpgt\ can be evaluated immediately
\eqn\kpr{G( p; \l , \l ') = \int _{0}^{\infty} dE
\langle \l | E \rangle {1 \over \cosh (\pi E) - \cos (\pi p)}
\langle E| \l ' \rangle}
 The quantum number  E plays the role of the momentum of an
additional "Liouville" dimension.

 The propagator  \kpr\  is universal in the sense that it
 does not depend on the background
momentum; it is the same for any closed string
theory with effective dimension (central charge
of the matter  in the language of $2d$ gravity) less than 1.

The poles of the propagator define the possible Liouville energies
corresponding to given momentum $p$
\eqn\lenr{i E_{n}(p)= \pm p + 2k , \ k \in \Z}
If we consider a theory with a background momentum
$p_{0}= |g-1|$
the allowed momenta are
$ p= n p_{0}, \ n \in \Z$.
For each value $p$ of the momentum together with the lowest energy states
 $E=\pm p$ (the creation and annihilation operators of a
 closed string
``tachyon'' ) there is an infinite discrete set of states which
describe infinitesimal deformations of the boundary at this point.
A boundary of finite length can be expanded as an infinite series of
such operators.

    Now let us calculate the  propagator for the open string.
A typical configuration of the loops on the world sheet
is shown in \fig\opr{A typical configuration of domain walls for the
open string propagator}. The generic loop configuration
involves three kinds of  domain walls: closed loops, lines
ending at the same boundary and lines connecting
 two different boundaries of the world sheet.
This last kind of domain walls describes the propagation of
the open string  in $x$-space. If we denote by
$\G_{0}(\l,\l')$ the amplitude of propagation between two consequent
domain walls with the lengths $\l$ and $\l '$, then the full propagator is
given by
\eqn\prpo{\G(p;\l , \l ' ) = \sum _{n=0} ^{\infty}
\int _{0}^{\infty} d\l _{1} ... d\l _{n}\Bigg({2\cos (\pi p)
\over 2\cos(\pi p_{0}/2)} \Bigg)^{n}
\G_{0}(\l ,\l _{1}) \G_{0}(\l_{1}, \l _{2}) ... \G_{0} (\l _{n} , \l ' ) }
i.e., by the same expression as \prpgt\ , with $G _{0}$ replaced by
$\G_{0}/\o $ and the measure $d\l /\l$ replaced by $d \l$.  The difference
between the two measures of integration is due to the cyclic symmetry
of the closed string
which is absent for the open string.
Now let us calculate $\G_{0}(\l , \l ')$.
 This is the loop amplitude for a disk with a boundary divided into four
 segments having alternatively
Dirichlet and Neumann boundary conditions.
Such an amplitude can be decomposed as a convolution of an amplitude
with Dirichlet boundary and a number of amplitudes with Dirichlet-Neumann
boundaries. The decomposition can be performed
 by cutting the world surface along
the most internal open lines.  Summing over the numbers $m$ and $n$ of
such lines at the two opposite boundaries we find
\eqn\convo{\eqalign{
\G_{0}(\l , \l ')=&
\sum _{k,m =0} ^{\infty} \int _{0}^{\infty} {\o \over  T}^{k+m}
\prod _{i=1}^{k} d\l _{i} e^{-2K_{0}\l _{i}} V(\l _{i})
 \cr &
\prod _{j=1}^{m} d \bar \l _{j} e^{-2K_{0} \bar \l _{j}} V(\bar \l _{j})
W(\l + \l ' + \sum _{i=1}^{k} \l _{i} + \sum _{j=1} ^{m} \bar \l _{j}) \cr
&=\sum _{m,k =0}^{\infty} \oint {dt \over 2\pi i} e^{t(\l + \l ')}
\W(t) [\V(2K_{0}-t)]^{k+m} [{\o \over  T}]^{k+m} \cr
 &= \oint {dt \over 2\pi i} e^{t(\l + \l ')}
 {\W(t) \over [1- \o  \V(2K_{0}-t) /T]^{2}} \cr}}
where product from 1 to 0 is assumed 1.
The contour of integration goes around the cut  $[t_{L},t_{R}]$ of
$\W_{0}(t)$, leaving outside the cut $[2K_{0}-t_{R}, 2K_{0}-t_{L}]$
of the denominator in the integrand.
Note that $\G_{0}$ depends only on the sum $\l + \l '$ of its arguments.

In the scaling limit we replace the quantities in the integrand
by their singular parts  according to eqs. \scprt\ and \sptw .
The regular part in the denominator cancels and we obtain the following
scaling limit of the irreducible part of the string propagator
\eqn\sclgr{ {1 \over \o}
\G_{0}(\l ,\l ') = \int _{M} ^{\infty} dz \ e^{-z(\l + \l ')}
\G_{0} (z)}
\eqn\sclgg{\G_{0}(z) = {1 \over \o}{{\rm Im }\w (-z) \over [\v(z)]^{2}}
={1 \over \o}{\i [\v(-z, \mu)] \over \v (z, \mu)} \sim 1 , \ \ \ z \to \infty}
(The asymptotic value at $z \to \infty$ follows
 from the large $z$ asymptotics of
the open string background amplitude
$\v(z)  \sim z^{g/2}$.

Before proceeding further let us notice that at  small  lengths
$\G_{0} (\l +\l')$  is identical to its analog  \eelcy\  for the
closed string, up to the factor $\sqrt{\l  \l '}$
\eqn\gam{{1 \over \o}\G_{0} (\l , \l ') = {e^{-M(\l +\l ')} \over \l +\l '}}
The factor $\sqrt{\l \l '}$ will appear if we replace the open string
integration measure $d\l = e^{\phi} d\phi$ with $d\l / \l = d\phi$.
Hence one can expect that the spectrum of the open string propagator is
the same as
the one
 of the closed string, since the spectrum
 is defined by the
small $l$ asymptotics of the propagator. However, the eigenfunctions
will be certainly different.

A rigorous proof of this statement can be done by demonstrating
that the traces of all powers of the two propagators are the same.

Let us
consider the trace of the open string  propagator \prpo         \
and perform the $\l$ integrations using the form \sclgr\ of $\G_{0}$.
The result is
\eqn\trce{\eqalign{
{\rm Tr} \G(p)
&= \int _{a} ^{\infty} d\l G(p; \l , \l )
\cr &=  \sum _{n= 1}^{\infty} [2\cos
 (\pi p)]^{n} \int _{M}^{1/a}...\int _{M}^{1/a}
dz_{1} ...dz_{n} { \G_{0}(z_{1}) ... \G_{0} (z_{n}) \over
(z_{1}+z_{2})(z_{2}+z_{3}) ... (z_{n}+z_{1})} \cr}}

The integral is logarithmically divergent for large $z$ and has to be
 cut off at
$z \sim 1/a$.
If we plug the large $z$  expansion
\eqn\expn{\G_{0}(z) = 1 + ({\rm const. }) z^{-g} + ...}
in \trce\ , the constant term reduces to exactly the same convolutive
integral as in the case of the closed string propagator. This term diverges
as  $\log (1/a)$. All subdominant terms will produce ${\it finite}$
corrections. Therefore
\eqn\blablabla{{\rm Tr} \G(p) = \lim _{a \to 0} \log ({1 \over a})
\int _{0}^{\infty} {dE \over \cosh (\pi E) - \cos (\pi p)}
\ \ \ + {\rm finite \ terms}}
One can repeat the same argument for any integer power of the propagator.
This means that the open string propagator can be written in a form similar to
\kpr
\eqn\kkpr{\G( p; \l , \l ') = \int _{0}^{\infty} dE
\langle \l | E \rangle _{\L , \mu} {1 \over \cosh (\pi E) - \cos (\pi p)}
\langle E | \l '\rangle  _{\L , \mu}}
All dependence on the cosmological constant $\L$ and the mass $\mu$ at the ends
of the string is absorbed in the eigenstates of the propagator
\eqn\eggst{\langle E|  \l \rangle _{\L , \mu}
=\langle \CV_{iE} \t  \CC (\mu)\CO(\l) \rangle _{\L}}
 Of course, the next terms of the asymptotics also contain some
universal information about the open string theory, and they will be
important for the calculation of the string interactions.
Note that
the on-mass-shell (microscopic) states $E=\pm ip$ are the order parameter
amplitudes \ampp .

Once the propagator is known, the amplitudes involving string interactions
can be calculated by decomposing the world sheet into irreducible
pieces (vertices and propagators)
in the same way as it has been done in the case of
the closed string. The dependence on the external momenta is only through
the propagators. For the three-string amplitude this is
illustrated by \fig\llast{Decomposition of the world sheet for the
three string amplitude}. The vertices $\G_{n} (\l_{1},\tl_{1},\l_{2},\tl_{2},
..,\l_{n+2},\tl_{n+2})$ represent amplitudes for a disk
with $n+2$ pairs of Neuman and Dirichlet boundaries with lengths $\l_{k}$
and $\tl_{k}$, $k=1,2,...,n+2$.
It is easy to see that these vertices will depend only on the  total lengths
$\l = \l_{1}+...+\l_{n+2} $ and $\tl =\tl_{1}+...+\tl_{n+2}$ of the
Dirichlet and Neumann boundaries.
Therefore the Laplace image of $\G_{n}$ (we denote it by the same letter)
will depend only on two variables $z$ and $\mu$. It is easy to establish the
following integral equation which follows from geometrical decomposition
of the world sheet shown in \fig\lasttt{The decomposition of the world sheet
producing the integral representation for the vertex $\G_{1}$}
\eqn\wwwu{\G_{n}(z, \mu)=
\int _{M}^{\infty}
{  {\rm Im} \w(-z) \over [\v( z, \mu )]^{n+2} }}
Therefore
\eqn\vrtyz{\i \G_{n}(z, \mu)= { {\rm Im} \w(z) \over [\v( -z, \mu )]^{n+2}}
\ \ \  {\rm along \ the\ cut\ } z<M}
In particular, the tadpole vertex  $(n=-1)$
is the basic open string
amplitude
\eqn\bassa{\G _{-1}(z,\mu)=\v(z, \mu)}
and the integral representation \wwwu\ becomes a closed equation
which is the continuum limit of \liii .

The spectrum of excitations of  the open string is fixed by the
set of allowed  target space momenta $p$ . Since the  background momentum
$p_{0}/2$
of the open string is twice less than that of the closed string,
 its spectrum will be twice denser
\eqn\spc{p= {1 \over 2} np_{0},  \ \ \ n \in \Z}

\newsec{Conclusion}

We have demonstrated in this paper an approach to the open string
theory which allow us not only to calculate the scaling dimensions of
the boundary operators, but also to obtain, in principle, any given
amplitude for the open strings in the dimensions less than 1, with
arbitrary in- and out- momenta. The propagator of two open strings
presented here is the simplest possible example. Technically, this
question is not simple, since one has to know the eigenfunctions of
this propagator, for which we know only the functional equation.

Another interesting possibility is to try to solve the whole field
theory for the open strings, which seems now to be a difficult but not
a hopeless task. This might shed some light on the string picture in
the multicoulour QCD according to an observation made in
\ref\KKKO{I.K.Kostov, \np 265 [FS15] (1986) 223    }.
This might be an object of further study.

Finally, let us note that the spectrum of the open string excitations is
not related to the value of the parameter $\mu$ which can be considered as
the mass of the ``quarks'' at the ends of the open string.

\bigbreak\bigskip\bigskip\centerline{{\bf Acknowledgements}}\nobreak

We thank F. David for his comments on the manuscript.

\appendix{A}{}

To give a more precise meaning to the derivation of the eq. \leqn
for the open string amplitude let us use its definition by means of a
three-positioned graphs
 as shown in the \fig\triang{Discretized world sheet
 of the open string amplitude  in the
 loop gas representation}. A world sheet of the
string amplitude $V _{\t m ,m}$ looks like a $\phi^{3}$
planar Feynman diagram with $\t m$ legs occupied by the ends of open
lines (thick  lines) and m nonoccupied legs ( thin  lines) at the boundary.
$\t m$ and $m$  are the lengths of the Neumann and Dirichlet boundaries,
 correspondingly. To every link on the graph occupied by a loop or open line
one subscribes a weight $1/(2K_{0})$.

If we pick up one leg at an edge of the Neumann boundary, the corresponding
 open line decomposes  a graph of the amplitude into two similar
 ones, only with the
different lengths of the Neumann and Dirichlet boundaries, and this
geometrically obvious decomposition allows us to write the loop equation in the
form:
\eqn\Ii{V_{\t m ,m}=  \sum _{k=2}^{\t m} \sum _{p,q=0}^{\infty} W_{k-2,q}
W_{\t m -k ,p+m} {(p+q)! \over p! q! } (2K_{0})^{-p-q-1}}
If we introduce the generating function
\eqn\Iii{\V(T,t)= \sum _{\t m ,m=0}^{\infty} T^{-\t m -1}
 t^{-m-1} V_{\t m ,m}}
this equation transforms in the following way

\eqn\iden{\eqalign{
&\V(T,t) -{1 \over T} V_{0}(t) =
 \sum _{\t m =2}^{\infty} T^{-\t m -1} \sum_{m=0}^{\infty} t^{-m-1}
\sum _{n=2}^{\t m}
 \oint {d\sigma \over 2\pi i }
 \oint {d\tau  \over 2\pi i } \cr
& V_{n-2}(\sigma) V_{\t m -n}(\tau)
\sum _{p,q=0}^{\infty}
{(p+q)! \over p! q! } (2K_{0})^{-p-q-1} \sigma^{q} \tau^{p+m} \cr}}
where $V_{k}(t)$ is the amplitude with $k$ legs on the Neumann boundary
($k$ is even, of course) as a
function of the spectral parameter $t$ of the Dirichlet boundary. In
particular, $V_{0}(t)=\W(t)$ is the amplitude with pure Dirichlet boundary.
Performing all the sums and the integration over $\sigma$ in \iden, we
arrive to the following equation

\eqn\Iiii{T \V(T, t)= \W(t) + \oint {d\tau \over 2\pi i}
{1 \over t-\tau}\V(T,\tau) \V(T, 2K_{0}-\tau)       }
which is identical (up to a normalization of $\V(T,t)$) to our main equation
\laplim.

%
%

\listrefs
\listfigs   
\bye